\documentclass[showpacs, preprintnumbers, prb, superscriptaddress, floatfix, twocolumn, longbibliography, footinbib]{revtex4-2}

\usepackage{amsmath, amsthm, amssymb}
\usepackage{graphicx}
\usepackage{xcolor}
\usepackage{xargs}
\usepackage[colorlinks,citecolor=blue,urlcolor=blue]{hyperref}

\usepackage[applemac]{inputenc}

\newcommand{\beq}{\begin{equation}}
\newcommand{\eeq}{\end{equation}}
\newcommand{\bea}{\begin{eqnarray}}
\newcommand{\eea}{\end{eqnarray}}

\newcommand{\mbt}{MnBi$_{2n}$Te$_{3n+1}$}
\newcommand{\mbttwo}{MnBi$_{2}$Te$_{4}$}
\newcommand{\mbtfour}{MnBi$_{4}$Te$_{7}$}
\newcommand{\mbteight}{MnBi$_{8}$Te$_{13}$}

\newcommand{\pd}{{\phantom\dag}}

\begin{document}

\title{Magnetic warping in topological insulators}

\author{Gabriele Naselli}
\affiliation{Institute for Theoretical Solid State Physics, IFW Dresden and W{\"u}rzburg-Dresden Cluster of Excellence ct.qmat, Helmholtzstr. 20, 01069 Dresden, Germany}

\author{Ali G. Moghaddam}
\affiliation{Department of Physics, Institute for Advanced Studies in Basic Sciences (IASBS), Zanjan 45137-66731, Iran}
\affiliation{Computational Physics Laboratory, Physics Unit, Faculty of Engineering and
Natural Sciences, Tampere University, P.O. Box 692, FI-33014 Tampere, Finland}
\affiliation{Institute for Theoretical Solid State Physics, IFW Dresden and W{\"u}rzburg-Dresden Cluster of Excellence ct.qmat, Helmholtzstr. 20, 01069 Dresden, Germany}

\author{Solange Di Napoli}
\affiliation{Departamento de F\'{i}sica de la Materia Condensada, GIyA-CNEA, Av. General Paz 1499, (1650) San Mart\'{i}n, Pcia. de Buenos Aires, Argentina}
\affiliation{Instituto de Nanociencia y Nanotecnolog\'{i}a (INN CNEA-CONICET), 1650 San Mart\'{i}n, Argentina}

\author{Ver\'onica Vildosola}
\affiliation{Departamento de F\'{i}sica de la Materia Condensada, GIyA-CNEA, Av. General Paz 1499, (1650) San Mart\'{i}n, Pcia. de Buenos Aires, Argentina}

\author{Ion Cosma Fulga}
\affiliation{Institute for Theoretical Solid State Physics, IFW Dresden and W{\"u}rzburg-Dresden Cluster of Excellence ct.qmat, Helmholtzstr. 20, 01069 Dresden, Germany}

\author{Jeroen van den Brink}
\affiliation{Institute for Theoretical Solid State Physics, IFW Dresden and W{\"u}rzburg-Dresden Cluster of Excellence ct.qmat, Helmholtzstr. 20, 01069 Dresden, Germany}

\author{Jorge I. Facio}
\affiliation{Institute for Theoretical Solid State Physics, IFW Dresden and W{\"u}rzburg-Dresden Cluster of Excellence ct.qmat, Helmholtzstr. 20, 01069 Dresden, Germany}
\affiliation{Instituto de Nanociencia y Nanotecnolog\'{i}a (INN CNEA-CONICET), 1650 San Mart\'{i}n, Argentina}
\affiliation{Centro At\'omico Bariloche and Instituto Balseiro, CNEA, 8400 Bariloche, Argentina}

\begin{abstract}
We analyze the electronic structure of topological surface states in the family of magnetic topological insulators \mbt. 
We show that, at natural-cleavage surfaces, the Dirac cone warping changes its symmetry from hexagonal to trigonal at the magnetic ordering temperature.
In particular, an energy splitting develops between the surface states of same band index but opposite surface momenta upon formation of the long-range magnetic order.
As a consequence, measurements of such energy splittings constitute a simple protocol to detect the magnetic ordering via the surface electronic structure, alternative to the detection of the surface magnetic gap. 
Interestingly, while the latter signals a nonzero surface magnetization, the trigonal warping predicted here is, in addition, sensitive to the direction of the surface magnetic flux. 
Our results may be particularly useful when the Dirac point is buried in the projection of the bulk states, caused by certain terminations of the crystal or in hole-doped systems, since in both situations the surface magnetic gap itself is not accessible in photoemission experiments.
\end{abstract}

\maketitle

\section{Introduction}

Antiferromagnetic topological insulators (AFM-TIs) are topological insulators that spontaneously break time-reversal symmetry ($\Theta$) while preserving the symmetry $S=\Theta T_{1/2}$, where $T_{1/2}$ is a lattice translation by half of a unit cell \cite{PhysRevB.81.245209}. 
The manifestations of the non-trivial topology on a given surface of such a system depend on whether or not the surface is symmetric under $S$. 
In the family \mbt, the crystal structure consists of septuple layers (SLs) of MnBi$_2$Te$_4$ separated by $n-1$ quintuple layers (QLs) of Bi$_2$Te$_3$ \cite{Otrokov2019, gong2019experimental}.
The Mn ions order in an uniaxial antiferromagnetic structure, with the N\'eel vector parallel to the stacking axis. 
The usually studied surfaces are terminated either on a QL or on a SL.
As a result, such surfaces are $S$-broken and in the magnetically ordered phase a gap $\Delta$ is allowed in the topological surface electronic spectrum.

$\Delta$ is expected to govern the low-energy physics when the chemical potential lies inside the gap, enabling phenomena such as the quantum anomalous Hall effect
\cite{PhysRevB.78.195424, PhysRevLett.102.146805, PhysRevB.92.081107, PhysRevB.93.045115, PhysRevLett.120.056801, mogi2017magnetic, Allen2019, liu2020robust, PhysRevLett.122.107202, lei2020magnetized, PhysRevB.103.235111}.
While the surface electronic structure has been extensively studied
\cite{Otrokov2019, gong2019experimental, PhysRevB.100.121104, PhysRevX.9.041038, PhysRevX.9.041039, PhysRevX.9.041040, Wueaax9989, Vidal2019, hu2020van, PhysRevB.101.161109, PhysRevB.101.161113, gordon2019strongly, PhysRevB.102.245136, Klimovskikh2020, xu2019persistent, PhysRevX.10.031013, PhysRevLett.126.176403, hu2020realization, PhysRevX.11.011039, shikin2021sample}, 
the observation and the temperature evolution of $\Delta$ remain as challenging and heavily debated issues. 
In the simplest picture, $\Delta$ should vanish at $T > T_N$ (above the N\'eel temperature) due to the restoration of $\Theta$ in a statistical sense.
Perhaps the clearest experimental data available today showing compelling evidence for this is found in the ferromagnetic compounds \mbteight\, \cite{PhysRevX.11.011039} and MnSb$_2$Te$_4$ ~\cite{wimmer2021mn} and in heterostructures  ~\cite{li2022large,liu2022visualizing,https://doi.org/10.48550/arxiv.2207.14421}. 
In \mbttwo, several photoemission experiments find within their resolution a gapless surface spectrum at all temperatures \cite{PhysRevX.9.041038,PhysRevX.9.041039,PhysRevX.9.041040}. 
In these cases, it has been argued that the surface magnetic structure might differ from the bulk one. 
On the other hand, experiments that do show a finite $\Delta$ at low temperatures do not always find it closes above $T_N$ \cite{Otrokov2019}.
Different possible mechanisms that might prevent the observation of the gap closing in photoemission experiments have been discussed, including short-range magnetic fields generated by chiral spin fluctuations \cite{Shikin2020}, anisotropy of the Mn-spin fluctuations \cite{PhysRevB.103.L180403}, and fermionic fluctuations \cite{PhysRevResearch.3.L032014}.
In addition, the hybridization between the Dirac cone and other trivial bands might further complicate the direct experimental study of $\Delta$.
This has been recognized as a problem at QL-terminated surfaces, where the Dirac cone has been theoretically and experimentally found to be buried in the surface projection of bulk bands, leaving at low energy the so-called hybridization gap \cite{PhysRevLett.126.176403}. 

Importantly, doping is ubiquitous in as-grown samples of \mbt, triggering the question of further consequences of breaking $\Theta$ associated with deviations from the low-energy Dirac limit.
In this paper, we analyze this issue and focus on the warping of the surface Dirac cone. 
In non-magnetic TIs, soon after their discovery it was observed that for usual levels of doping, the surface Fermi contour presents a pronounced hexagonal warping \cite{chen2009experimental, xia2009observation, PhysRevLett.103.146401} and the importance of $\Theta$ being preserved for the resulting hexagonal symmetry was recognized early on \cite{PhysRevLett.103.266801}. 
Here, we show that the opening of a magnetic gap $\Delta$ in $S$-broken surfaces is generically accompanied by a reduction in the symmetry of the Dirac cone warping from hexagonal to trigonal. 
As we will show, $\Delta$ also plays a role in the low-temperature trigonal warping of the Dirac cone. 

From these results, a simple protocol emerges to detect the long-range magnetic order via the surface electronic structure: measurements of the energy difference between states of same band but of opposite surface momenta should vanish in the paramagnetic phase and become finite in the ordered phase. 
This protocol may prove particularly useful on terminations of a crystal where the Dirac point is known to be buried in the projection of the bulk states or in $p$-doped systems, since both conditions make the magnetic gap $\Delta$ inaccessible to photoemission experiments. 
In addition, the trigonal distortion exhibited by the surface Dirac cone warping in the low-temperature phase is sensitive not only to the existence of a net surface magnetic flux but also to its direction.

This article is organized as follows. 
In Sec.~\ref{2x2}, we analyze the problem based on general symmetry considerations for a $2\times2$ model Hamiltonian of the (001) surface Dirac cone. 
In Sec.~\ref{4x4}, we construct a tight-binding model Hamiltonian for AFM-TIs having three-fold rotational symmetry with respect to the N\'eel propagation vector of the AFM phase together with reflection symmetries with respect to planes containing such axis. 
This model allows us to continuously vary the sublattice magnetization mimicking the evolution of temperature in uniaxial antiferromagnets. 
In Sec.~\ref{dft}, we present ab-initio results of finite slabs of \mbtfour. 
In Sec.~\ref{sec_discussion} we discuss experimental aspects of our predictions.  
Finally, Sec.~\ref{conclusions} provides concluding remarks.

\section{Effective $2\times2$ Hamiltonian for the surface Dirac cone}
\label{2x2}

We start by exploring the symmetries and surface electronic properties of AFM-TIs based on an effective two-band Hamiltonian for their topological surface states.

\subsection{Symmetries}

\begin{figure}[t]
 \centering
 \includegraphics[width=8.5cm]{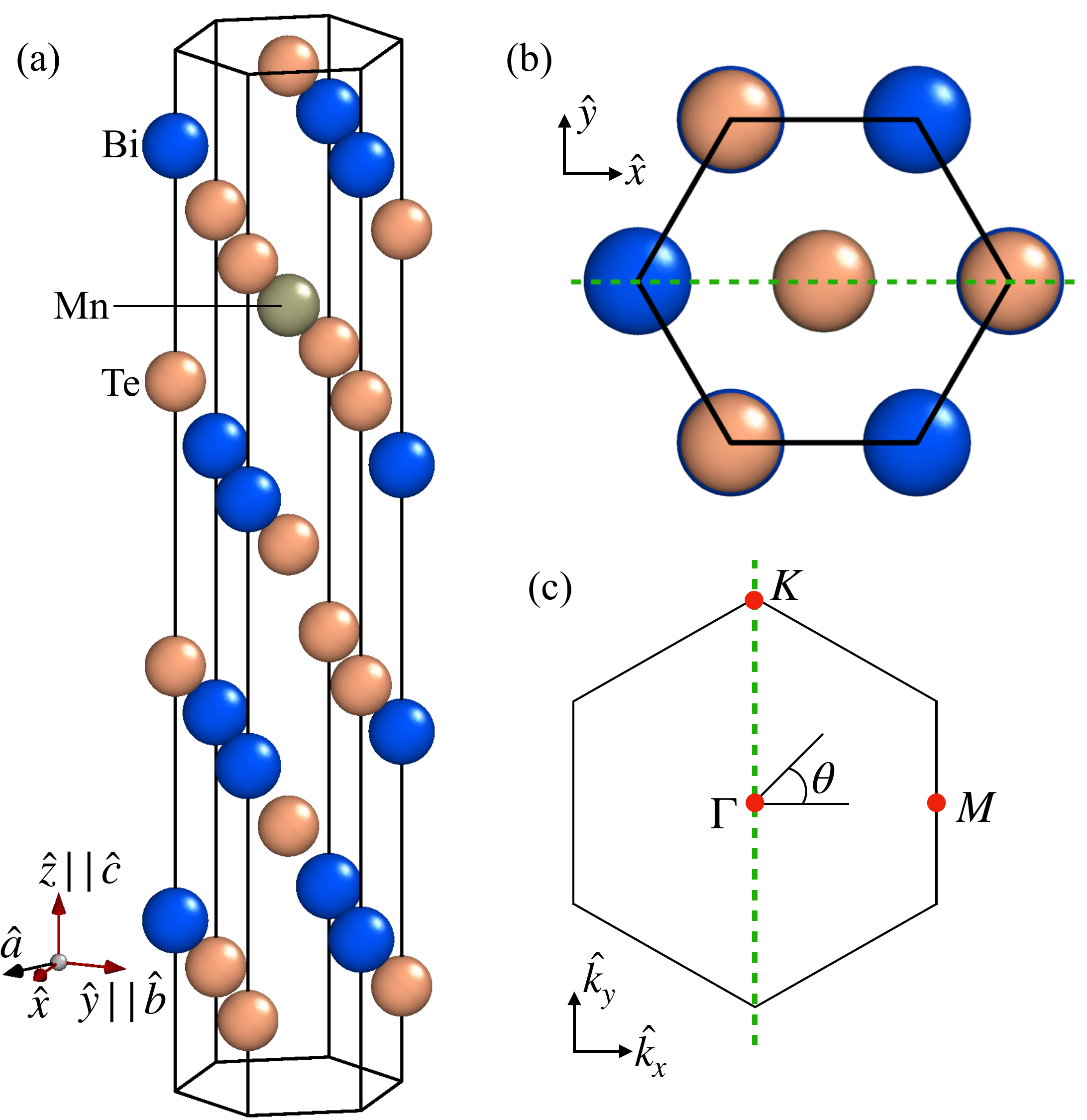}
	\caption{ 
	(a) Crystal structure of MnBi$_4$Te$_7$. 
	(b) Top view of the crystal structure.
	(c) Surface Brillouin zone.
	The green dashed line in (b) indicates the reflection symmetry plane $M: \hat{y} \to -\hat{y}$ and in (c) the high-symmetry plane corresponding to the combined symmetry $M^\prime = M \Theta$, with $\Theta$ the time-reversal operator.
	}
	\label{struct}
\end{figure}

The natural-cleavage planes of compounds in the family \mbt\, are positioned at by Te-Te van der Waals gaps, here considered to be perpendicular to the $\hat{z}$ axis.
The case $n=2$ is illustrated in Fig.~\ref{struct}.
The crystal structure is minimally described either with the space group $R3\bar{m}$ (when $n$ is even) or with $P\bar{3}m1$ (odd $n$). 
In both cases, the surface crystal point symmetry is $C_{3v}$, which can be generated by a three-fold rotation $C_{3z}$ and a reflection symmetry $M: y \to -y$. 

When including spin-orbit coupling, the magnetic point symmetry depends on the direction of the Mn magnetic moments.
We will consider the case typically observed in bulk magnetometry experiments, where Mn magnetic moments point along the trigonal axis \cite{zeugner2019chemical,PhysRevMaterials.3.064202,PhysRevLett.124.197201}. 
In this case, the electronic structure is symmetric under $C_{3z}$ and under the antiunitary reflection symmetry $M^\prime = M \Theta$.
Notice that the momenta invariant under $M$ obey $k_y = 0$ while those invariant under $M^\prime$ satisfy $k_x=0$.

Throughout this section, in order to construct $2\times2$ Hamiltonians for the surface Dirac cone, we use the representations
\begin{subequations}
\begin{align}
	C_{3z} &= \exp (-i\sigma_z \pi/3), \\
	M &= i \sigma_y, \\
	\Theta &=i\sigma_y K, \\
	M^\prime &= -K
\end{align}
\end{subequations}
where $\sigma_i$ are the Pauli matrices and $K$ is complex conjugation.
When a Hamiltonian $H$ is said to be symmetric under $C_{3z}$, $M$, $\Theta$, or $M^\prime$, it satisfies
\begin{subequations}
\label{eq_allsym}
\begin{align}
	C_3^{-1} H(k_x,k_y) C_3 &= H\Big(\frac{-k_x+\sqrt{3}k_y}{2},\frac{-\sqrt{3}k_x-k_y}{2}\Big) \label{eq_C3}\\
	M^{-1} H(k_x,k_y) M &= H(k_x,-k_y), \label{eq_mirrorM} \\
	\Theta^{-1} H(k_x,k_y) \Theta &= H(-k_x,-k_y), \label{eq_TR} \\ 
	{M^{\prime}}^{-1} H(k_x,k_y) M^\prime &= H(-k_x,k_y), \label{eq_mirrorA}
\end{align}
\end{subequations}
respectively.

\subsection{Time-reversal symmetric case}

We begin by revisiting the results for non-magnetic topological insulators of the family Bi$_2$Te$_3$ derived in Ref.~\cite{PhysRevLett.103.266801} and consider the Hamiltonian
\begin{equation}
	H_{\Theta}(\mathbf{k}) = v (k_y \sigma_x-k_x \sigma_y) + i\frac{\lambda}{2}(k_{+}^3-k_{-}^3)\sigma_z
	\label{eq_H_TR}
\end{equation}
where $\mathbf{k}=(k_x, k_y)$ is the momentum vector, $k=|\mathbf{k}|$, and $k_\pm=k_x\pm ik_y$. 
This Hamiltonian obeys Eqs.~\eqref{eq_allsym}.
Defining the azimuth angle $\theta$ with respect to the $\hat{x}$ axis, the energies of the bands are
\begin{align}
	&\varepsilon_{\pm}(\mathbf{k}) = \pm \sqrt{ v^2k^2 + \lambda^2k^6\sin^2(3\theta)} ,
\end{align}
Warping effects originate from the term proportional to $\lambda$ and the Fermi contour has a hexagonal symmetry, $\theta \to \theta + 2\pi/6$, with the vertices of the hexagon pointing along $\Gamma M$.

\subsection{Hexagonal warping as a helical correction to the magnetic gap}

Consider adding to Eq.~\eqref{eq_H_TR} the simplest $\Theta$-breaking term
\begin{equation}
H_\Delta(\mathbf{k}) = H_{\Theta}(\mathbf{k}) + \Delta \sigma_z.
	\label{eq_H_Delta}
\end{equation}
This Hamiltonian obeys Eq.~\eqref{eq_C3} and breaks both Eqs.~\eqref{eq_mirrorM} and \eqref{eq_TR}, but satisfies their combination, Eq.~\eqref{eq_mirrorA}. 
The energies are
\begin{align}
	&\varepsilon_{\pm}(\mathbf{k}) = \pm \sqrt{\Delta^2 + v^2k^2 - 2 \lambda \Delta k^3 \sin(3\theta) + \lambda^2k^6\sin^2(3\theta)}.
	\label{eq_e_delta}
\end{align}
In this case, the spectrum acquires a gap $\Delta$ and the symmetry of the Fermi contour is lowered to trigonal $\theta \to \theta + 2\pi/3$ due to the term proportional to $\sin(3\theta)$. 

\begin{figure}[t]
 \includegraphics[width=8.5 cm]{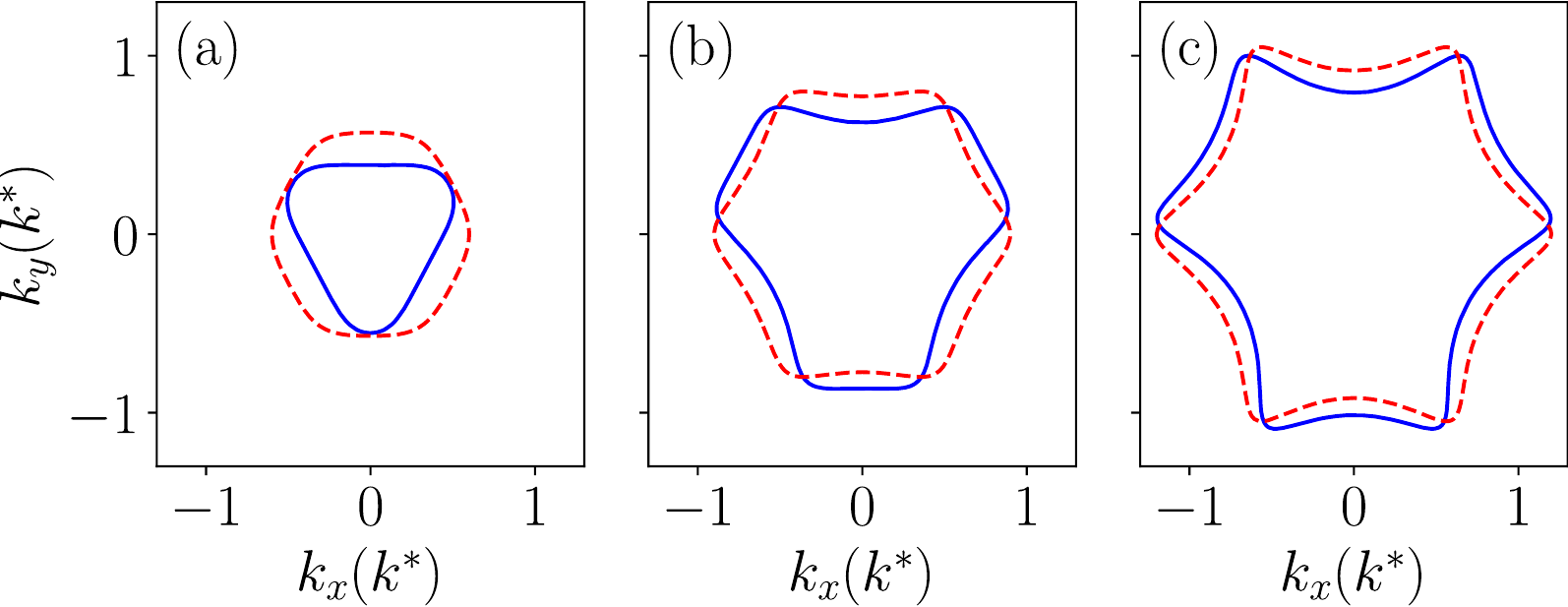}
	\caption{
	(a-c) Fermi contour for Fermi energies $E_F/E^*=0.6$,  $0.9$ and $1.2$. 
	Dashed red and solid blue curves correspond to $\Theta$-symmetric ($\Delta/E^*=0$) and $\Theta$-broken ($\Delta/E^*=0.4$) cases, respectively.
    } 
	\label{fig_FS}
\end{figure}

Figure \ref{fig_FS}(a-c) shows the Fermi contour for Fermi energies $E_F/E^*=0.6$, $0.9$, and $1.2$, respectively, where we have used the units of energy $E^*=v/k^*$ and of momentum $k^*=\sqrt{v/\lambda}$, as introduced in Ref.~\cite{PhysRevLett.103.266801}. 
The latter is a natural unit to account for the length scale associated with the hexagonal warping.
For each Fermi energy, the Fermi contours corresponding to $\Delta/E^*=0$ and $0.4$ are shown. 
As a reference, notice that for the case $\Delta=0$, the parameters used in panel (c) match those found in Ref.~\cite{PhysRevLett.103.266801} to reproduce experimental results in Sb$_2$Te$_3$ ($v_0=2.55$\,eV\AA, $\lambda=250$\,eV\AA$^3$, $E^*=0.23$\,eV and $E_F=1.2 E^*$).
For these parameters, $\Delta=0.4E^* \sim 0.1\,$eV. 
While the trigonal distortion is apparent in all cases, it becomes less pronounced for relatively large doping because of the larger power in $k$ associated with the hexagonal warping term in Eq.~\eqref{eq_e_delta}.

\begin{figure}[t]
 \includegraphics[width=7.8 cm]{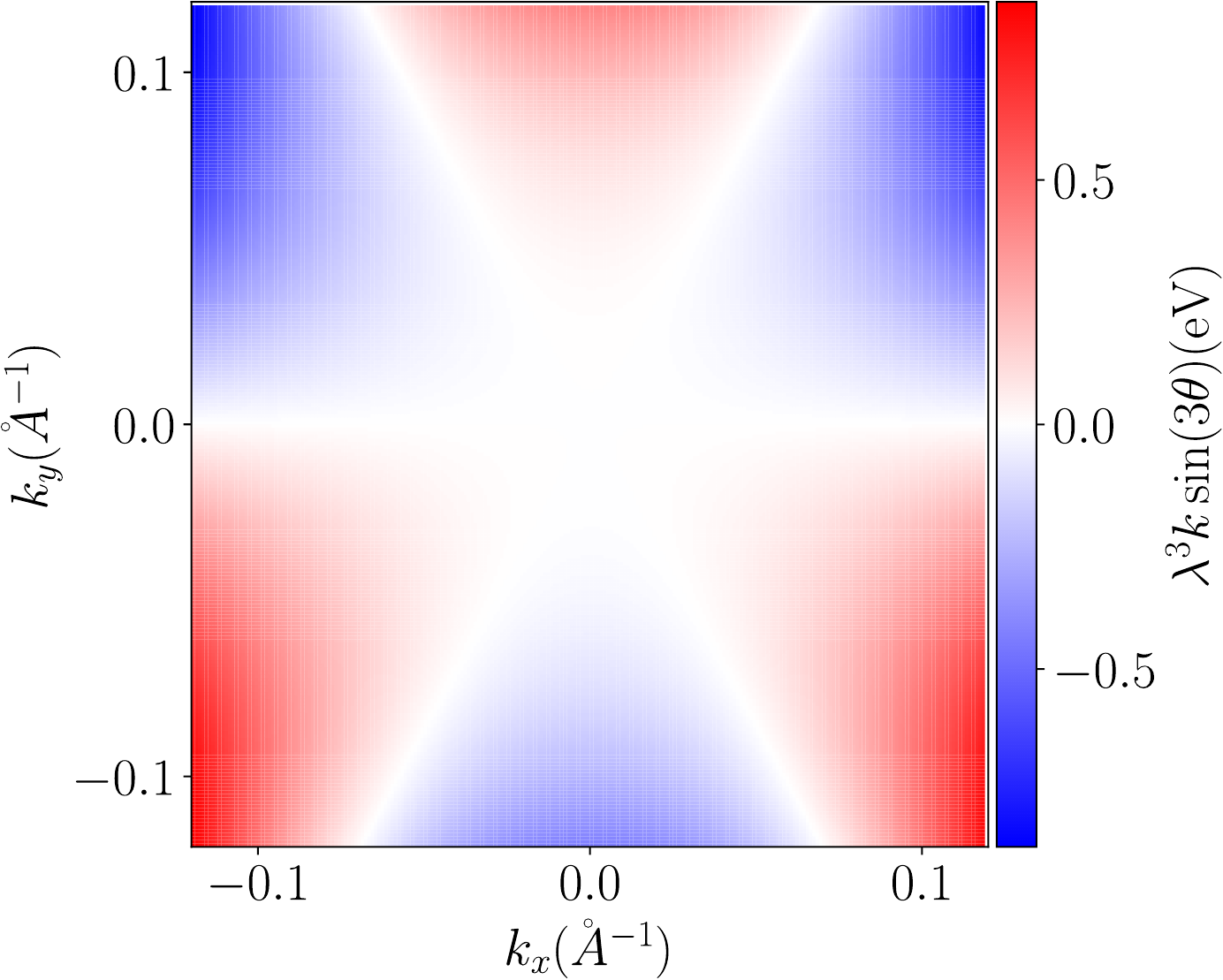}
	\caption{
	Momentum dependence of the hexagonal warping term in eV for $\lambda=250$\,eV\AA$^3$.
	}
	\label{fig_warping}
\end{figure}

\begin{figure}[t]
 \includegraphics[width=8.5 cm]{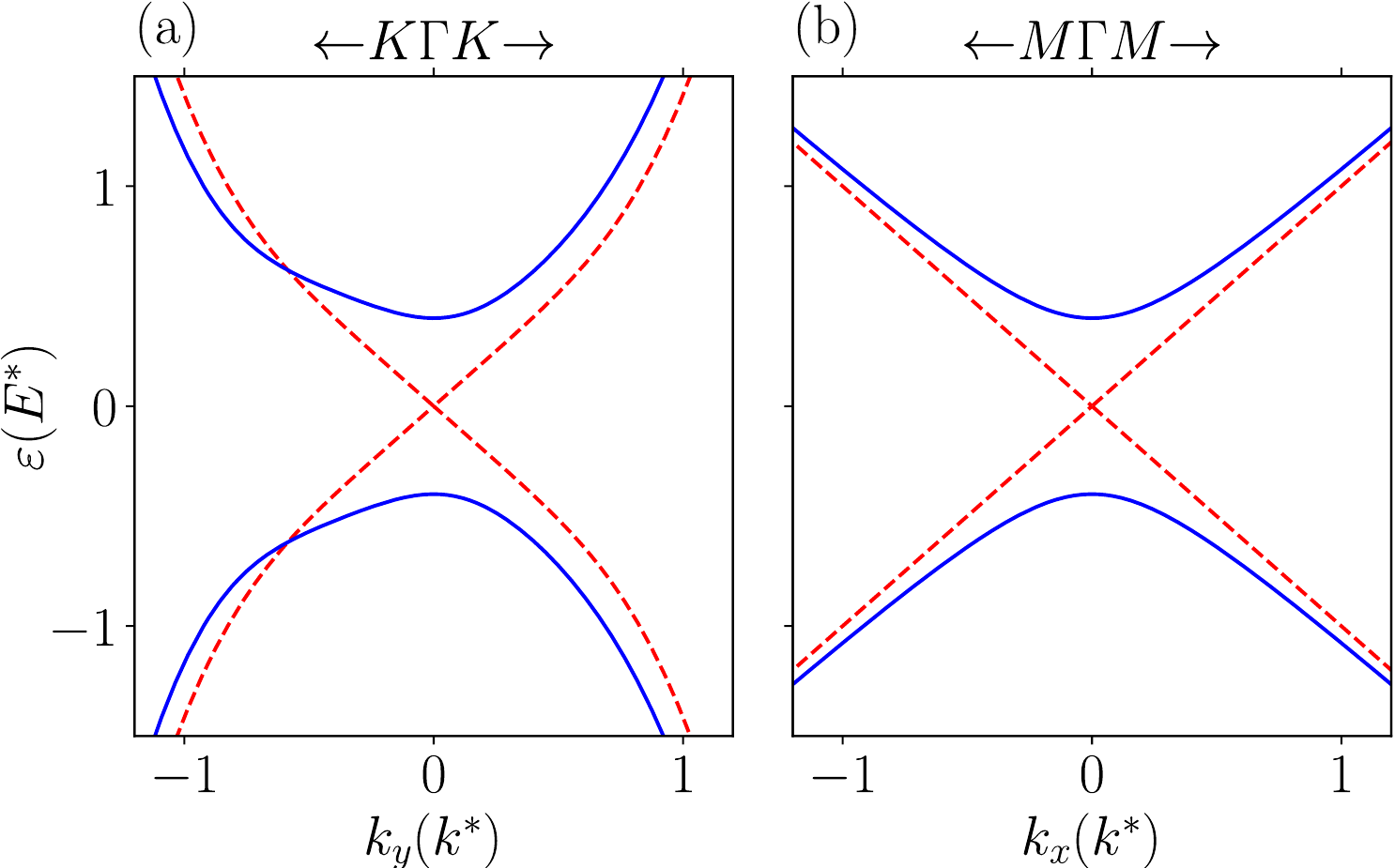}
	\caption{
	Energy disperson for the model Eq.~\eqref{eq_H_Delta} along $\Gamma K$ (a) or $\Gamma M$ (b).
	Dashed red and solid blue curves correspond to $\Theta$-symmetric ($\Delta/E^*=0$) and $\Theta$-broken ($\Delta/E^*=0.4$) cases, respectively.
	}
	\label{fig_dirac}
\end{figure}

Within this model, the product of $\Delta$ and $\lambda$ characterizes the strength of the trigonal distortion.
This can be traced back to the fact that two terms associated with these parameters are proportional to the Pauli matrix $\sigma_z$. 
Therefore, one can define an effective gap function 
\begin{equation}
	\Delta_{\rm eff}(k,\theta) = \Delta + \lambda k^3 \sin(3\theta).
	\label{eq_Deltak}
\end{equation}
While the first term induces a net spin moment and breaks $\Theta$, the second term induces a helical, momentum-dependent spin texture compatible with $\Theta$.
Figure \ref{fig_warping} shows this term using the value of $\lambda=250$\,eV\AA$^3$, which was found in Ref.~\cite{PhysRevLett.103.266801} to appropriately describe the hexagonal warping in Sb$_2$Te$_3$.
This term cannot by itself open a gap at $\Gamma$, where it vanishes.
For finite $\Delta$, however, it does affect the energy dispersion of the gapped Dirac cone in a quite distinctive manner.
Taking, for example, momenta along the $M^\prime$-invariant line $k_x=0$ ($\Gamma K$), for $v k_y \gg \Delta $ the eigenenergies of the Hamiltonian Eq.~\eqref{eq_H_Delta}, linearized in $\Delta$ and $\lambda$, read 
\begin{equation}
	\pm v |k_y| \Big(1 + \frac{\Delta \lambda}{v^2} k_y\Big).
\end{equation}
Thus, for finite $\Delta$, the hexagonal warping term bends the Dirac cone. 
This bending is forbidden along $\Gamma M$ due to Eq.~\eqref{eq_mirrorA} (see Fig.~\ref{fig_dirac}).

\subsection{General case}

Using Qsymm \cite{varjas2018qsymm}, we find that the most general Hamiltonian symmetric under $C_{3z}$ and $M^\prime$ up to third order in momentum is
\begin{equation}
	H(\mathbf{k}) = E_0(\mathbf{k}) + H_\Delta(\mathbf{k})
	+ \beta \left( \begin{matrix}
                     0       & -k_{+}^2 \\
                     k_{-}^2 & 0
	               \end{matrix} \right),
	\label{eq_Hkp}
\end{equation}
where
\begin{equation}
	E_0(\mathbf{k}) = \big[ \frac{k^2}{2m^*} - i\frac{\gamma}{2} (k^3_{+}-k^3_{-}) \big] \sigma_0.
\end{equation}
Notice that, in general, the Fermi velocity $v$ and the magnetic gap $\Delta$ appearing in $H_\Delta(\mathbf{k})$ admit quadratic corrections
\begin{align}
	  v(k) &= v (1+ \nu k^2), \\
	  \Delta(k) &= \Delta (1+ \delta k^2), 
\end{align}
respectively. The Hamiltonian in Eq.~\eqref{eq_Hkp} satisfies symmetries given by Eqs.~\eqref{eq_C3} and \eqref{eq_mirrorA}, and its energy bands read
\begin{align}
	&\varepsilon_{\pm}(k,\theta) = \frac{k^2}{2m^*} + k^3 \gamma \sin(3\theta) \pm \Big[ \Delta^2 + v^2(k)k^2 - \nonumber \\ 
	& 2 [\beta v(k)+\lambda \Delta(k)] k^3 \sin(3\theta) + \beta^2 k^4 + \lambda^2k^6\sin^2(3\theta) \Big]^{1/2}. 
\end{align}
The terms that break $\Theta$ are those associated with $\beta$, $\gamma$, and $\Delta$, and all of these contribute to the trigonal distortion. 
While $\beta v(k)$ affects the warping in a similar way to the product $\lambda \Delta(k)$, the term proportional to $\gamma$ is diagonal in spin subspace and has a different power in momenta.

\section{Effective lattice Hamiltonian for magnetic TI\lowercase{s} with $C_{3v}$}
\label{4x4}

The models studied in the previous section contain the basic ingredients to describe the hexagonal to trigonal transition of the Dirac cone warping, but have a fundamental shortcoming that various parameters associated with the breaking of $\Theta$ are generally allowed.
Their relative importance is not obvious without additional information, e.g., from experiment. 
In this section, we analyze the bulk electronic structure and topological surface states based on tight-binding models that obey the point symmetries of different compounds in the \mbt\ family.

We first consider a model for a strong topological insulator (STI) on a triangular lattice with Bravais vectors: 
\begin{align}
         \begin{bmatrix}
           {\bf a}_1 \\
           {\bf a}_2 \\
           {\bf a}_3
         \end{bmatrix}
	 &=
	 \begin{bmatrix}
		 1 &0& 0 \\
		 -1/2 &\sqrt{3}/2& 0 \\
		 0 &0& 1
         \end{bmatrix}
	 \begin{bmatrix}
		 \hat{x} \\
		 \hat{y} \\
		 \hat{z}
         \end{bmatrix}.
  \end{align}
We define the crystal vector momentum ${\bf k} = (k_1, k_2, k_3)$, with $k_j = {\bf k}\cdot {\bf a}_j$ and consider the Bloch Hamiltonian:
\begin{equation}\label{eq:HTB}
\begin{split}
	H_{\rm TI}({\bf k}) = & \Gamma_1 [ \mu +f(k_1,k_2) - \cos(k_3)] \\
 & +\lambda^\pd [ \Gamma_2 \sin(k_1) + \Gamma_{2,1} \sin(k_2) \\
 & \qquad - \Gamma_{2,2} \sin(k_1+k_2) + \Gamma_3 \sin(k_3) ].
\end{split}
\end{equation}
The scalar function $f(k_1,k_2)$ is:
\begin{equation}
	f(k_1,k_2) = -\sum_{j=0}^{2} \cos({\cal C}_3^j k_1), \label{eq:fG}\\
\end{equation}
and $\Gamma_1=\tau_z\sigma_0$, $\Gamma_2=\tau_x\sigma_x$, and $\Gamma_3=\tau_y\sigma_0$, where Pauli matrices $\tau$ encode the degree of freedom associated with two orbitals per site.
${\cal C}_3$ denotes a threefold rotation applied to the momentum, such that $k_1 \to k_2$, $k_2 \to -k_1-k_2$, and $-k_1-k_2 \to k_1$.
The matrices $\Gamma_{2,1}$ and $\Gamma_{2,2}$ are related to $\Gamma_2$ by a threefold rotation $C_3$:
\begin{equation}\label{eq:rotGamma}
\Gamma_{2,j} = C_3^j \Gamma_2 C_3^{-j}, \quad C_3=\tau_0 \exp \left( i\frac{\pi}{3}\sigma_z \right).
\end{equation}

The Hamiltonian in Eq.~\eqref{eq:HTB} has inversion symmetry $I=\tau_z\sigma_0$, time-reversal symmetry $\Theta=i\tau_0\sigma_yK$, three-fold rotation symmetry $C_3$ 
and reflection symmetry $M$ with respect to the $k_1=-2k_2$ plane (and analog planes related by $C_3$). 
Setting $\mu=3$ and $\lambda=1$, we obtain an STI with $\mathbb{Z}_2$ indices $(\nu_0;\nu_1\nu_2\nu_3)=(1;000)$ and a mirror-protected topological crystalline insulator \cite{PhysRevMaterials.3.074202}.

\begin{figure}[h]
 \centering
 \includegraphics[width=8.5cm]{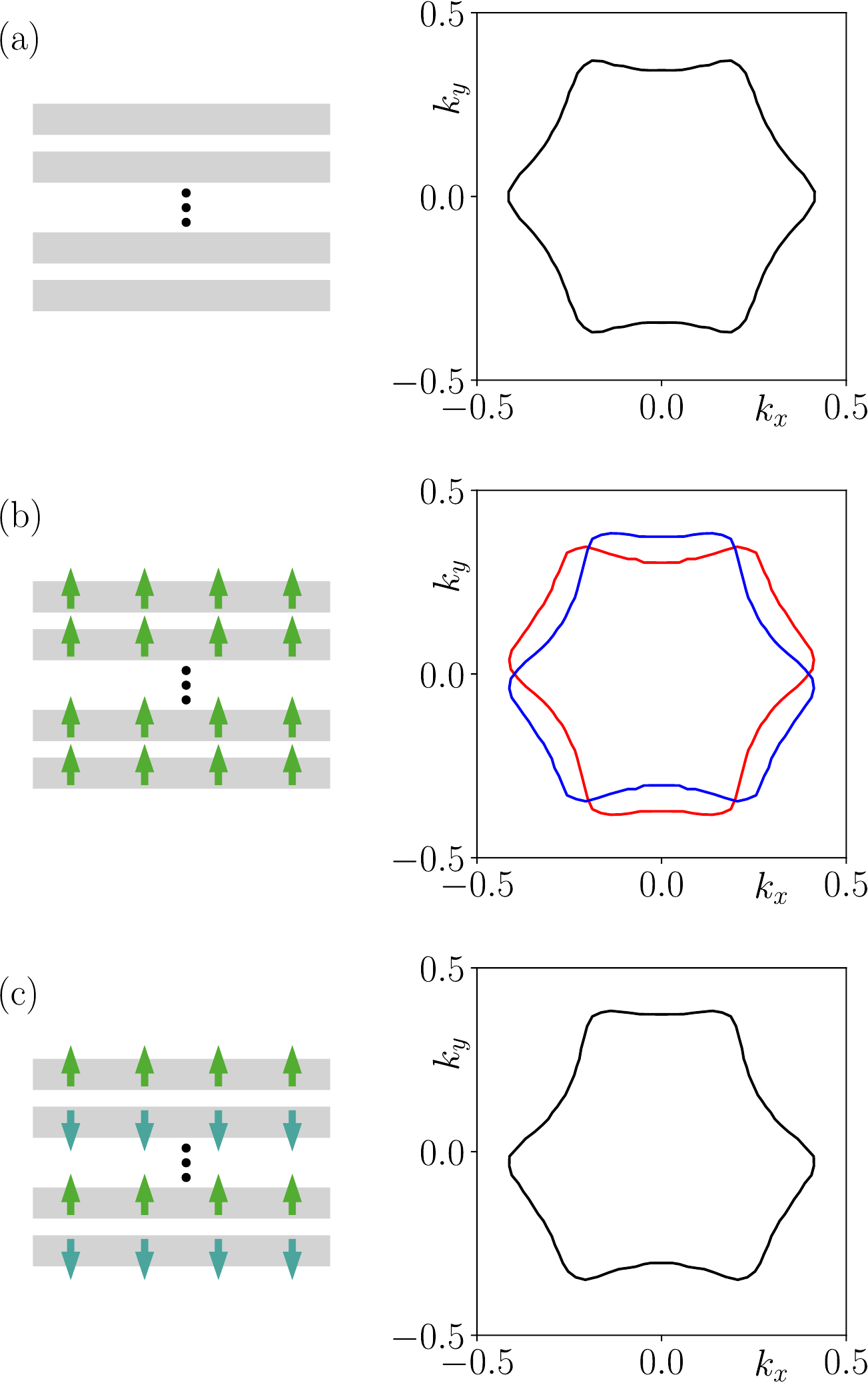}
	\caption{
	(a) Surface Fermi contour for a tight-binding Hamiltonian of a strong topological insulator with $C_{3v}$ symmetry. 
	The iso-energetic contour has a sixfold rotational symmetry. 
	(b) Surface Fermi contour in the presence of a Zeeman field. 
	The Fermi contour of the top and bottom surfaces are rotated by $\pi/3$ with respect to each other and both exhibit a trigonal symmetry. 
	Each band has only a trigonal symmetry due to the time-reversal symmetry breaking. Bands associated with states at the bottom and top surfaces are colored in red and blue respectively.
	(c) Surface Fermi contour for an antiferromagnetic topological insulator. 
	For an even number of magnetic layers, the surface magnetic flux at opposite surfaces is identical and so are the trigonally-distorted Fermi contours. 
	Parameters were chosen as $\mu=3$, $\lambda=1$, $t=10$, $b=0.2$, and Fermi energy $E=0.7$ for a system with $20$ layers stacked along the $z$ direction.
	}
	\label{fermi_surface_tight}
\end{figure}

For the surface states to have a pronounced hexagonal warping in the $\Theta$-symmetric phase, as usually observed experimentally, we add a next-nearest-neighbor intralayer hopping term
\begin{equation}\label{eq:nextnearest}\\
\begin{split}
	H_{t}({\bf k})=&t\Gamma_t[\sin(k_1+2k_2)+\\
	&\sin(k_1-k_2)-\sin(2k_1+k_2)], 
\end{split}
\end{equation}
with $\Gamma_t=\tau_x\sigma_z$. 
This term obeys time-reversal $\Theta$, three-fold rotation $C_3$ and the mirror symmetries. 
Figure \ref{fermi_surface_tight}(a) shows the surface Fermi contour obtained by solving the Hamiltonian $H_{\rm TI}+H_t$ in a slab geometry consisting of a finite number of layers in the $\hat{z}$ direction.
For the calculations in this section we use the Kwant code \cite{Groth_2014}. 

We now consider the breaking of the time-reversal symmetry via a Zeeman field:
\begin{equation}
	H_b=b S_z, \label{eq:Zeeman}\\
\end{equation}
where $S_z=\tau_0\sigma_z$. 
This term breaks both the time-reversal and mirror symmetry but preserves their product, $\Theta M$.
Figure \ref{fermi_surface_tight}(b) shows the surface Fermi contour with this added perturbation. 
Compared to the $\Theta$-symmetric case, the warping exhibits a trigonal symmetry. 
In addition, the Fermi contours at opposite surfaces of the slab (shown in red and blue) are rotated with respect to each other. 
This rotation originates from the fact that, at opposite surfaces of the finite slab, the field $H_b$ points in opposite directions relative to the surface normal. 
Thus, the surface states at the two surfaces experience an opposite magnetic flux.

We now build a tight-binding model appropriate for an AFM-TI.
We follow the procedure introduced in Ref.~\cite{PhysRevB.81.245209}, which consists in adding a staggered $\Theta$-breaking term to a Hamiltonian that describes an STI. 
To do this, we double the unit-cell long the $\hat{z}$ direction and add a sublattice degree of freedom represented by the Pauli matrices $\gamma_i$ ($i=0,\dots 3$). 

We introduce the Hamiltonian
\begin{equation}
	H_{\rm AFM}=[H_{\rm TI}({\bf k})+ H_t( {\bf k})]\gamma_0 + bS_{z} \gamma_z. \label{eq:antiferro}\\
\end{equation}
In this model the Zeeman coupling is opposite for adjacent layers. 
Importantly, this staggered field is the only term associated with the breaking $\Theta$, and it related to the average Mn sublattice magnetization in a transparent way. 
 
Figure \ref{fermi_surface_tight}(c) shows the corresponding surface Fermi contour. 
For an even number of magnetic layers, the surface magnetic flux at opposite surfaces points along the same direction relative to the surface normal.
Accordingly, in this case the Fermi contour at both surfaces exhibits a trigonal distortion of identical shape.

Both the results for the FM and for the AFM models show that, on a given surface, the relative orientation of the trigonal distortion is essentially set by the local direction of the surface magnetization with respect to the surface normal. As a consequence, if there exist pieces of a given surface where the surface magnetization points in opposite directions, our results suggest that the trigonal distortions observed in both cases should be oriented differently and related by a $\pi/3$ rotation.
This phenomenology can be directly appreciated in the simplest model Hamiltonian Eq.~\eqref{eq_H_Delta}. 
Without hexagonal warping ($\lambda=0$), the band structure [Eq.~\eqref{eq_e_delta}] depends only on $\Delta^2$ so that information on the phase of $\Delta$ is lost. 
This is no longer the case for finite $\lambda$, for which the energy bands depend on the sign of $\Delta \lambda$. 
As a consequence, assuming a value of $\lambda$ independent of temperature and $\Delta \propto M_s$, with $M_s$ the surface magnetization, the trigonal distortion in the warping yields information on the sign of $M_s$ (or, equivalently, of $\Delta$).

\section{Ab initio results}
\label{dft}

\begin{figure}[t]
 \centering
 \includegraphics[width=8.5cm]{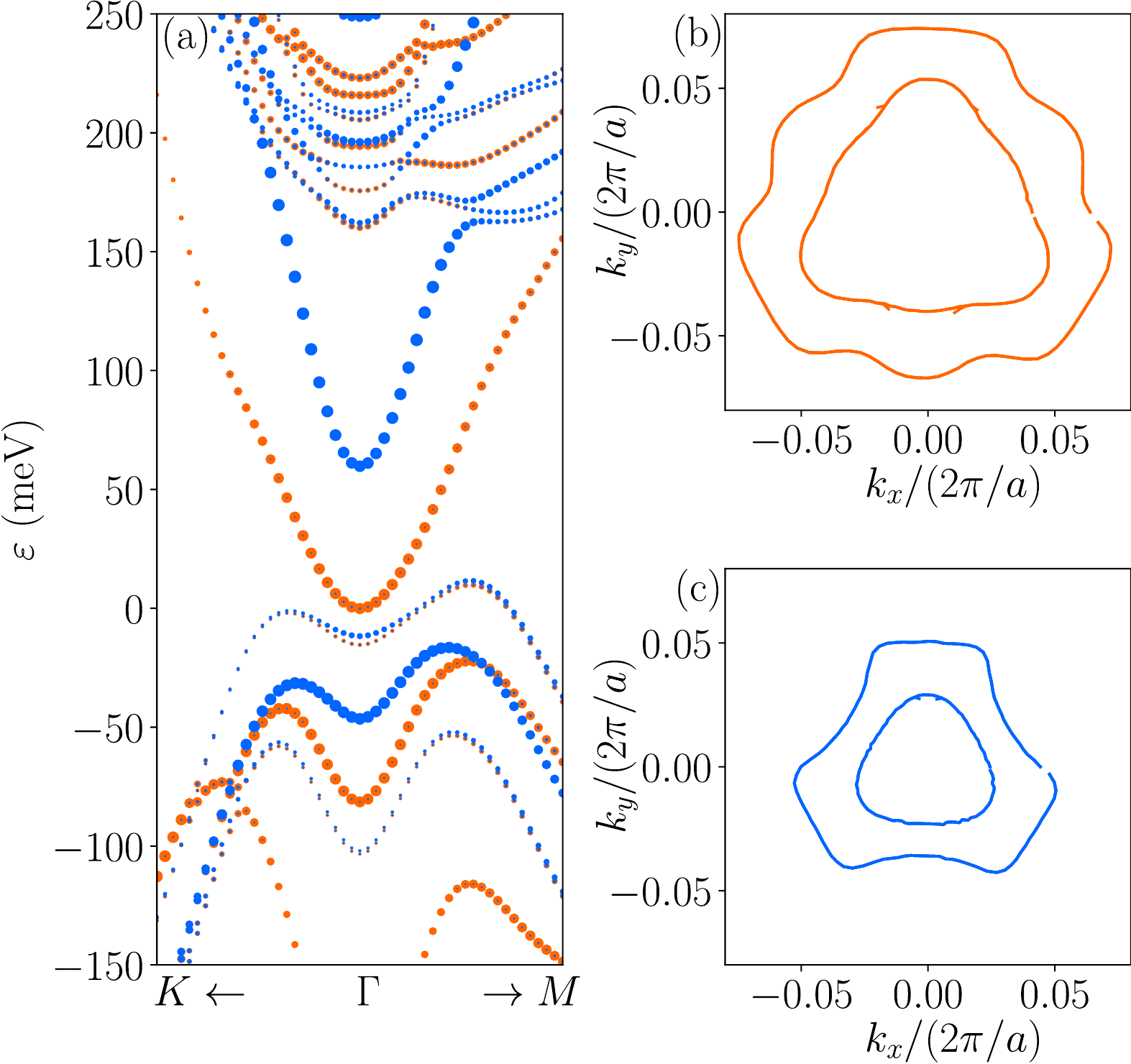}
	\caption{
	(a) MnBi$_4$Te$_7$ band structure projected on the outermost quintuple layer (orange) and on the outermost septuple layer (blue). 
	Energies are measured with respect to the Fermi energy of the full slab.
	(b) Constant energy contours of quintuple layer-projected band-structure at energies 100 and 150\,meV. 
	(c) Same as panel (b) for the septuple layer termination.
	}
	\label{dft_147}
\end{figure}

In this section, we present density functional theory (DFT) calculations based on finite slabs of \mbtfour.
We use a structural model consisting of four unit cells and of a vacuum of 30 Bohr radii.
We use the experimental bulk lattice parameters and atomic positions.
The calculations are based on the GGA$+U$ method with the generalized gradient approximation \cite{PhysRevLett.77.3865} as implemented in the FPLO code version 48.00-52 \cite{PhysRevB.59.1743}. 
We fix parameters $U=5.34\,$eV and $J=0$, as in Ref.~\cite{Otrokov2019}, and use the atomic limit flavor for the double counting correction.
The spin-orbit interaction is considered in the fully-relativistic, four-component formalism. 
Numerical $k$-space integrations are performed with a triangle method with a mesh of $12\times12\times1$ subdivisions in the Brillouin zone.

Figure \ref{dft_147}(a) shows the band-structure of \mbtfour\, projected on the outermost SL and on the outermost QL of the slab.
The dependence on the surface termination has been studied in detail experimentally and theoretically in Ref.~\cite{PhysRevLett.126.176403}.
One important point is that for the SL termination the apparent gap is actually a magnetic gap, while for QL-termination the magnetic gap is buried into the projection of bulk bands.
Figures \ref{dft_147}(b) and (c) show constant-energy contours which evidence the trigonal symmetry of the warping, regardless of the surface termination. 

While here we only present results for \mbtfour, theoretically-obtained surface Fermi contours with a substantial trigonal warping is also for \mbttwo, e.g. in Ref.~\cite{PhysRevB.101.161109}.

\section{Discussion}
\label{sec_discussion}

The results presented above follow from the observation that long-range magnetic order will generally reduce the symmetry of the topological surface states, effectively introducing a warping in their Fermi contours. 
While we have focused on the magnetic structure that has been experimentally determined from several bulk magnetometry techniques, it is always conceivable that other surface phenomena induce modifications in the surface magnetic structure. 
For example, in-plane magnetic moments at the surface have been suggested to play a role in the thermal evolution of the surface gap (e.g. Ref.~\cite{PhysRevX.9.041038}). 
In such cases, the magnetic warping of the surface states could also reveal useful information. 
For example, if the magnetic moments lie in-plane and remain collinear, the Fermi contours should reflect the breaking of the three-fold rotational symmetry. If, additionally, the magnetic moments point perpendicular to one of the three crystal reflection-symmetry planes, the Fermi contours should exhibit a reflection symmetry with respect to that plane.

In photoemission experiments, an important aspect which should be taken into account is that the photoemission intensity is not solely determined by the single-particle spectral function ${\cal A}({\bf k},\omega)$, but it also depends on the photoemission matrix elements which incorporate the transition between the initial and final states \cite{lv2019ARPES}.
These can provide information on the orbital characters of electronic states \cite{Rotenberg2008graphene, yi2011superconductor}, but may also obscure the characterization of anisotropies of the band energy dispersion \cite{Lanzara2011spinARPES, eremeev2012spinARPES, Herdt2013spinARPES}.
In particular, matrix elements can introduce an asymmetry in the measured spectral weight of states with opposite in-plane momenta. Nevertheless, using a variable photon energy and also changing the orientation of the setup with respect to the crystal can make it possible to discern bulk from surface states and probe the inherent warping of the latter \cite{chen2009experimental, Ando2014warping}.

Another possibly relevant technical aspect is the presence of twin domains, which can in principle mix Fermi contours with their $\pi/3$-rotated partners.
While their presence can be detrimental to the experimental detection of the trigonal warping, their density can be reduced with adjusted growth methods \cite{Cryst.Growth, li2010van}. 

The magnetic warping could also be probed by using scanning tunneling microscopy to measure quasiparticle interference (QPI) patterns around surface impurities \cite{Zhang2009QPI, beidenkopf2011spatial}. As already shown for time-reversal symmetric TIs, the almost parallel segments of hexagonally warped isoenergy lines provide a nesting condition, producing highly focused QPI patterns with similar six-fold symmetric Fourier transforms \cite{Sessi2016,Thalmeier2020}.
For magnetic topological insulators with trigonally warped energy contours, we expect the nesting to be reduced, leading to the appearance of a three-fold symmetric QPI pattern. 
Lastly, the change in the symmetry of the warping may become an important ingredient in various surface phenomena associated with Dirac physics, such us surface magnetic textures \cite{PhysRevResearch.3.033156, PhysRevResearch.3.033173, PhysRevB.105.035156} and anomalous surface transport properties \cite{PhysRevB.102.205407, PhysRevB.103.035410, PhysRevB.103.075424}.

\section{Conclusions}
\label{conclusions}

We have analyzed how long-range magnetic order affects the symmetry of the surface Dirac cone in the family of magnetic topological insulators \mbt.
With this aim, we have introduced effective continuum as well as tight-binding models obeying the relevant symmetries of the problem.
Based on these models and on density-functional calculations in the ordered phase, we have shown that, in addition to opening a gap $\Delta$ in the surface electronic structure, the magnetic order lowers the symmetry of the Dirac cone warping from hexagonal to trigonal. 
As a consequence, surface states of opposite momenta (or, in general, of states of surface momenta related by a $n \pi/3$ rotation, with $n$ odd) become energy-split below the ordering temperature, as recently noticed also in related preprints~\cite{Binghai2022, Wang2022}. 
This observation may provide a practical protocol to detect the magnetic transition via the surface electronic structure, particularly useful in cases where the magnetic gap $\Delta$ is inaccessible to photoemission experiments either because the Dirac cone is buried into the projection of bulk bands or because the system is hole-doped.
Furthermore, as opposed to the surface magnetic gap, the trigonal distortion is sensitive to the sign of the surface magnetization.

\section{Acknowledgments}

We thank Anna Isaeva, Hendrick Bentmann and Eike F. Schwier for fruitful discussions, Ulrike Nitzsche for technical assistance, and the DFG for support through the W\"urzburg-Dresden Cluster of Excellence on Complexity and Topology in Quantum Matter, ct.qmat (EXC 2147, project-id 390858490) and through SFB 1143 (project-id 247310070) project A5. 
J.I.F. would like to thank the support from the Alexander von Humboldt Foundation and ANPCyT grants PICT 2018/01509 and PICT 2019/00371

\bibliography{ref}

\end{document}